\documentstyle[12pt,aps,epsf]{revtex}
\begin{document}
\draft
\tighten
\title{
$J/\psi$ normal and anomalous suppressions in a hadron and string
cascade model}
\vspace{0.1in}
\author{
Sa Ben-Hao,$^{1,2,4}$ ~
Amand Faessler,$^{1}$ ~
Tai An,$^{3}$ ~
T. Waindzoch,$^{1}$ ~
C. Fuchs,$^{1}$ ~
Z. S. Wang,$^{1}$ ~ and
Wang Hui$^2$ ~
}
\vspace{0.1in}
\address{
$^1$ Institute of Theoretical Physics, University of Tuebingen,
  Auf der Morgenstelle 14, D72076 Tuebingen, Germany \footnotemark \\
$^2$ China Institute of Atomic Energy, P.O. Box 275 (18),
  Beijing, 102413 China  \\
$^3$ Institute of High Energy Physics, Academic Sinica,
   P.O. Box 918, Beijing, 100039 China\\
$^4$ Institute of Theoretical Physics, Academic Sinica,
   Beijing, China
\footnotetext{ Mailing address.}
}
\date{\today}
\maketitle
\vspace{0.1in}
\begin{abstract}
A mechanism for the effective dissociation of a $c\bar{c}$ pair in the colour
electric field of strings is introduced into a hadron and string cascade
model, i.e. JPCIAE, which is based on the LUND model, simulating ultra -
relativistic nucleus - nucleus collisions. This new mechanism together with
the known mechanism of nuclear absorption (both baryons and mesons) could
reproduce fairly the data of the normal and anomalous $J/\psi$ suppressions in 
minimum bias pA, AB (with light projectile), and Pb + Pb collisions at 200 A 
GeV/c. However the impact parameter (E$_T$) dependence of the $J/\psi$ 
suppression factor, both, in S + U and Pb + Pb reactions at 200 A GeV/c and 
158 A GeV/c, respectively, is not well reproduced. We also tested the 
additional mechanism of the energy degradation of leading particles, with 
which both, the normal and anomalous $J/\psi$ suppressions in minimum bias pA, 
AB, and Pb + Pb collisions and the E$_T$ dependence of the $J/\psi$ 
suppression factor are better reproduced.
\end{abstract}
\vspace{0.1in}

\section{Introduction}
It was suggested by Matsui and Satz [1] that the suppression of the $J/\psi$ 
yield in relativistic nucleus - nucleus collisions might be a powerful 
signature for the Quark Gluon Plasma (QGP) formation. Since that time, a 
number of corresponding experiments have been stimulated [2 - 4] to measure 
the $J/\psi$ yield via its dimuon decay. A significant suppression of the 
$J/\psi$ yield from pA collisions to AB collisions has already been observed 
in these experiments. However, except the anomalous suppression observed in 
Pb + Pb reactions at 158 A GeV/c [4] the normal suppression in pA and AB 
collisions with light projectiles has been well explained within the hadronic 
regime [5 - 6] either by Glauber theory (absorption model) [7 - 16] or by 
transport (cascade) model simulations [17 - 20]. The mechanism of the 
anomalous $J/\psi$ suppression in Pb + Pb collisions is, however, still a 
question of current debate [10 - 16, 18, 21 - 25].

Refs. [22] and [24] claim that the anomalous suppression can be explained
within the Glauber approach by introducing the comover absorption. However, in
Ref [23] one author of [22] has made a reexamination of the comover effect and
indicated that the comover mechanism ``cannot explain the observed $J/\psi$
suppression in Pb + Pb interactions". On the other hand, in Refs. [11 - 14,
16, 25] different extra suppression mechanisms due to the formation of the
QGP have been introduced to explain the anomalous $J/\psi$ suppression in
Pb + Pb collisions.

Recently in [26] the mechanism of the dissociation of $c\bar{c}$ pairs in
the colour electric field of a string as proposed in [27 - 28] has been 
included in a transport approach by a purely geometrical way. In [26] it is 
claimed that both, the data showing normal and anomalous $J/\psi$ suppression,
can be reproduced without assuming the formation of QGP in these collisions.

In this paper a mechanism for the effective dissociation of $c\bar{c}$ pairs
in the colour electric field of strings is introduced into the hadron and
string cascade model, i. e. JPCIAE [19], which is based on the LUND model
and the PYTHIA event generator [29]. This new mechanism together
with the already existing mechanisms, i. e. the nuclear absorption (both by 
baryons and mesons) could explain fairly both, the normal and anomalous 
$J/\psi$ suppression in minimum bias pA, AB, and Pb + Pb collisions at 200 A 
GeV energy. However the impact parameter (E$_T$) dependence of the $J/\psi$ 
suppression factor both in S + U and Pb + Pb reactions at 200 A GeV/c and 158 
A GeV/c, respectively, is not well reproduced. We investigated also 
the additional mechanism of the energy degradation of leading particles. 
Including this effect we obtain a better description of both, the normal
and anomalous $J/\psi$ suppressions in minimum bias pA, AB, and Pb + Pb
collisions. In particular, the E$_T$ dependence of the $J/\psi$ suppression 
factor is better reproduced by reducing the effective cross sections and the
model parameter concerning to the colour dissociation mechanism.
\section{Brief description of the model}
In JPCIAE the colliding nucleus is depicted as a sphere with radius $\sim$
1.05 A$^{1/3}$ fm ($A$ refers to the atomic mass number of a nucleus) in
its rest frame. The spatial distribution of nucleons in this frame is sampled
randomly due to a Woods - Saxon distribution. The projectile nucleons are
assumed to have an incident momentum per nucleon and the target nucleons are
at rest. That means the Fermi motion in a nucleus and the mean field of a
nuclear system are here neglected due to relativistic energy in question. For
the spatial distribution of the projectile nucleons the Lorentz contraction is
taken into account. A formation time is given to each particle and a particle
starts to scatter with others after it is ``born''. The formation time is a
sensitive parameter in this model.

A collision time is calculated according to the requirement that the minimum
approaching distance of a colliding pair should be less or equal to the value
$\sqrt{\sigma_{tot}/\pi}$, where $\sigma_{tot}$ is the total cross section of
the colliding pair [30 - 32, 19]. The minimum distance is calculated in the
CMS frame of the two colliding particles. If these two particles are moving
towards each other at the time when both of them are ``born'' the minimum
distance is defined as the distance perpendicular to the momenta of both
particles. If the two particles are moving back-to-back the minimum distance
is defined as the distance at the moment when both of them are ``born''. All
the possible collision pairs are then ordered into a collision time sequence,
called the collision time list. The initial collision time list is composed of
the colliding nucleon pairs, in each pair here one partner is from the
projectile nucleus and the other from the target nucleus.

Then the pair with the least collision time in the initial collision time list
is selected to start the first collision. If the CMS energy, $\sqrt s$, of this
colliding pair (a hadron - hadron collision) is larger than or equal to $\sim$
4 GeV two string states are formed and PYTHIA (with default parameters) is
called to produce the final state hadrons (scattered states). Otherwise no 
string state is formed and the conventional scattering process (elastic and  
inelastic scatterings or resonance production) [30 - 32, 19] is executed. 
After the scattering of this colliding pair, both the particle list and the 
collision time list are then updated and they are now not only composed of the
projectile and target nucleons but also the produced hadrons. Repeat the
previous steps to perform the second collision, the third collision, $\cdots$,
until the collision time list is empty, i.e. no more collision occurs in the
system.

In our model the $J/\psi$ is produced via QCD hard processe
\begin{equation}
g + g \rightarrow J/\psi + g.
\end{equation}
In PYTHIA a large number of QCD parton - parton processes have been
considered, including the $J/\psi$ production, Eq. (1). A user is allowed to
run the program with any desired subset of these processes. Thus, when
$\sqrt{s}$ is larger or equal $\sim$ 10 GeV the PYTHIA generator with the
selection of $J/\psi$ production is called. The produced particles as well
as the $J/\psi$ are then transported in the Monte Carlo simulation in the same
way as the nucleons. Any operation of desired processes, including
the $J/\psi$ production channel defined here, is a kind of bias sampling,
which enhances the probabilities of those desired processes. In order to
overcome the corresponding bias, the execution of PYTHIA with 
the $J/\psi$ production channel is additionally weighted with a 
production probability. This probability is equal to
the parameterized $J/\psi$ hadronic production cross section [33]
\begin{equation}
\sigma_{NN\rightarrow J/\psi +X} = d(1-\frac{c}{\sqrt {s}})^{12} ,
\end{equation}
(with c=3.097 GeV, d=2.37/B$_{\mu \mu}$ nb, and B$_{\mu \mu}$ = 0.0597
is the branching ratio of a $J/\psi$ to dimuon) multiplied by a factor. That
factor is adjusted such that the number of $J/\psi$ produced in each
event is around one, the same as in the experiments.

One point which should also be mentioned here is the fact that in the original 
JETSET program, which deals with the fragmentation of a string and runs 
together with PYTHIA, the leading particle in a nucleon - nucleon collision is 
assumed to carry about half of the incident energy. But the experiments of 
proton - nucleus and nucleus - nucleus collisions at relativistic energies 
reveal that, on the average, the produced nucleon carries only a smaller 
fraction of initial energy in each binary nucleon - nucleon collision [34]. A 
stopping law has been proposed in [34 - 36] to handle this situation. We have 
also applied this stopping law to calculate the energy fraction which the 
leading particle keeps after each binary nucleon - nucleon collision.

\section{Effective dissociation of a $c\bar{c}$ state in the colour electric 
field of string}
In [19] we have demonstrated that the hadron and string cascade model (usual
scenario simulation in [19]) could reproduce all the NA38 pA and AB (with light
projectile) data but could not reproduce the NA50 data of Pb + Pb collisions
even the effective cross sections ($\sigma_{J/\psi-B}^{Abs}$ = 6 mb and
$\sigma_{J/\psi-M}^{Abs}$ = 3 mb, where the subscripts B and M refer to baryon
and meson, respectively) have been enlarged by twenty percent. Thus in order to
understand the NA50 data of Pb + Pb collisions one has to invoke some new
mechanism which is more or less absent in pA and AB (with light projectile)
collisions but plays a role in large systems as Pb + Pb.

As discussed in [37] the object which is followed on its way out of hadronic
medium is not a $J/\psi$ but a meson ``in the making", which is called a
``premeson" or a ``pre-resonant state". The exact nature of the premeson is
still debated [37 - 38]. However, in the early stage of the nucleus - nucleus
collision this premeson or $c\bar{c}$ state should be surrounded by strings
[26] and once the $c\bar{c}$ state gets into the colour electric field of a
string it might be dissociated [26 - 27]. As a first step towards this
scenario, we treat this dissociation mechanism before the formation of a
physical $J/\psi$ in an effective way.

One knows that this dissociation effect increases with an increasing energy 
density stored in the strings, which, in turn, would be increase with the 
reaction energy, the centrality, and the size of reaction system [39]. Since 
the production and dissociation of $c\bar{c}$ state are mainly hard processes, 
the size dependence should follow an ``AB scaling" [4, 23, 26] (A (B) stands 
for the atomic mass number of projectile (target) nucleus, respectively). The 
centrality (impact parameter) dependence of this effective dissociation is 
modelled by a Gaussian distribution, in common use [24 - 25, 40].

Thus, the effective dissociation probability of a $J/\psi$ premeson,
$c\bar{c}$ state, is assumed to be
\begin{eqnarray}
P_d = \frac{\rho_s - \rho_s^{min}}{\rho_s^{max} - \rho_s^{min}}, \\
\rho_s = a_s\sqrt{s_{NN}}f(b)AB, \\
f(b) = \exp(-\frac{b^2}{R_L^2}).
\end{eqnarray}
In these equations $\rho_s$ refers to the energy density stored in the
strings, $\rho_s^{min}$ and $\rho_s^{max}$ are the corresponding lower and
upper limits : $P_d$ = 0 when $\rho_s$ = $\rho_s^{min}$ and $P_d$ = 1 when
$\rho_s$ = $\rho_s^{max}$. $\sqrt{s_{NN}}$ is the CMS energy of initial
nucleon - nucleon system.  R$_L$ refers to the larger value between the R$_A$
and R$_B$ (radii of projectile and target nuclei, respectively). Since it is
reasonable to assume $\rho_s^{min}$ = 0, the value of $\rho_s^{max}$ can
be absorbed into the model parameter and $P_d$ is reduced to
\begin{equation}
P_d = c_s\sqrt{s_{NN}}f(b)AB.
\end{equation}

The $c\bar{c}$ state created by calling PYTHIA has now the probability (1 -
$P_d$) to survive, it is regarded as a physical $J/\psi$ after formation time
and joined into the transport processes together with hadrons. Doing this, the
physical $J/\psi$ may interact with baryons and mesons which is described by
the effective cross sections [7 - 14, 41]: $\sigma_{J/\psi-B}^{abs}$ = 6 mb
and $\sigma_{J/\psi-M}^{abs}$ = 3 mb.  The formation time of meson and 
$J/\psi$ are assumed, as the same as in [19], to be $\tau_M$ = 1.2 fm/c and 
$\tau_{J/\psi}$ = 0.6 fm/c, respectively. The following reactions of a 
$J/\psi$ with baryons and mesons [17 - 19, 38] are considered
\begin{eqnarray}
J/\psi + N \rightarrow \Lambda_c + \bar{D} ,\\
J/\psi + \pi \rightarrow D + \bar{D^*}, \\
J/\psi + \rho \rightarrow D + \bar{D}.
\end{eqnarray}

The experimental $J/\psi$ suppression factor is defined as [4, 41]
\begin{equation}
S_{exp.}^{J/\psi} = (\frac{B_{\mu\mu}\sigma_{AB}^{J/\psi}}{AB})
                    /(B_{\mu\mu}\sigma_{pp}^{J/\psi}),
\end{equation}
and the theoretical $J/\psi$ suppression factor is expressed as [18 - 19, 41]
\begin{equation}
S_{theo.}^{J/\psi} = \frac{M_{J/\psi}}{M_{J/\psi}(0)}
\end{equation}
where $M_{J/\psi}(0)$ refers to the multiplicity of the primary $J/\psi$
generated by PYTHIA and $M_{J/\psi}$ to the multiplicity of $J/\psi$'s which
actually have survived the dissociation and the final state interactions.

Fig. 1 compares the $J/\psi$ suppression factors in minimum bias pA, AB, and 
Pb + Pb collisions at 200 A GeV/c as a function of AB to the corresponding 
NA38 and NA50 data. In the calculations the total cross sections are assumed 
to be $\sigma_{J/\psi-B}^{tot}$ = 8.64 mb and $\sigma_{J/\psi-M}^{tot}$ = 4.80 
mb which correspond to twenty percent enlarged values of $\sigma_{J/\psi-B}^
{abs}$ = 6 mb and $\sigma_{J/\psi-M}^{abs}$ = 3 mb above. The model parameter 
$c_s$ in Eq. (6)  is treated as a fit parameter which is mainly adopted to the 
Pb+Pb datum point. It takes the value 1.*10$^{-6}$. The open circles with 
error bars in Fig. 1 indicate the preliminary experimental data (cited 
directly from [41]). In Fig. 1 the dashed and dash - dotted curves are the 
theoretical results of the calculations with abs. + dis. (absorption + 
dissociation) and with abs. only. One sees from Fig. 1 that the effective 
dissociation mechanism of a $c\bar{c}$ pair in the colour electric field of 
strings does not contribute in pA and AB (with light projectile) collisions 
but plays a significant role in Pb + Pb collision which allows the 
reproduction of NA50 data.

In Figs. 2 and 3 the $J/\psi$ suppression factor is shown as a function of
the neutral transverse energy $E_T$ (GeV)
\begin{equation}
E_T = \sum_i E_i \times sin(\theta_i)
\end{equation}
(where E$_i$ and $\theta_i$ are the total energy and the polar angle of
neutral particle i, respectively) for S + U reactions at 200 A GeV/c and Pb + 
Pb reactions at 158 A GeV/c, respectively. The total cross section used in 
calculations are $\sigma_{J/\psi-B}^{tot}$ = 8.64 mb and $\sigma_{J/\psi-M}^
{tot}$ = 4.80 mb. The full squares with error bars in Fig. 2 and the full 
squares and circles with error bars in Fig. 3, respectively, refer to the 
preliminary NA38 and NA50 data (cited directly from [26]). The different 
curves in these figures correspond to the different calculations as explained 
in Fig. 1. Figs. 2 and 3 show that the experimentally observed $E_T$ 
dependence of the $J/\psi$ suppression factor is not well reproduced by the 
theoretical results, in particular in the Pb + Pb reaction. It should, however, 
be mentioned that the calculated transverse energy $E_T$ in Pb + Pb collisions 
is less than experimentally observed. In [42] it is mentioned: `` ... recently 
the E$_T$ scale has been recalibrated to be $\sim$ 10 - 20 smaller than shown 
here, ... the data situation remain unclear for both, the low and the high 
E$_T$ limit, ... ". Taking this reduction into account the calculated 
transverse energy $E_T$ is still a little lower than the data. This could be 
attributed to the fact that the transverse flow is not described in JPCIAE. 
There is, however, experimental evidence for a strong transverse flow in 
central Pb + Pb collisions [43].

\section{Test of the mechanism of leading particle energy degradation}
The energy degradation of leading particles (mainly nucleons) refers to the
following: If a projectile nucleon collides sequentially with two target
nucleons, the energy of projectile nucleon in the second collision is
reduced comparing to incident energy. Thus the corresponding probability of
the $J/\psi$ production in the second collision should be reduced as well
according to the parameterised $J/\psi$ hadronic production cross section,
Eq. (2).

Presently, the influence of this mechanism in $J/\psi$ suppression is a 
strongly debating issue. One might assume that this mechanism could lead to a 
violation of the AB scaling of the Drell - Yan production. However, there is 
really a violation of a strict AB scaling in $J/\psi$ production, since after 
rescaling the NA38 and NA50 data from different energies to 200 A GeV all the 
data except Pb + Pb lie on the universal curve (AB)$^{0.91}\times\sigma^
{J/\psi}_{pp}$. Thus, one should not rule out the effect of leading particles 
energy degradation on the $J/\psi$ suppression before more detailed further 
investigations.

Indeed, in [44] it was proved that the nuclear Drell - Yan ratio is suppressed 
because of the energy loss of leading particle. In [45] it has been declared 
that `` Precisely measured Drell - Yan cross sections for 800 GeV protons 
incident on a variety of nuclear targets exhibit a deviation from linear 
scaling in the atomic number A. ... this deviation can be accounted for by 
energy degradation of the proton as it passes through the nucleus if account 
is taken of the time delay of particle production due to quantum coherence.". 
It is also mentioned in [42] that ``Quantum effects such as energy dependent 
formation and coherence lengths must be taken into account before definite 
statements can be made with regard to the nature of the $J/\psi$ suppression."
. It is worthy, however, to test the role played by the mechanism of leading 
particle energy degradation in the $J/\psi$ suppression.

We repeat the calculations of Fig. 1 to 3 with the additional mechanism of
leading particle energy degradation and the results are given in Fig. 4 to 6, 
respectively. The effective cross sections used in the calculations are reduced 
to $\sigma_{J/\psi-B}^{tot}$ = 7.2 mb and $\sigma_{J/\psi-M}^{tot}$ = 4.0 mb (
which are corresponding to $\sigma_{J/\psi-B}^{abs}$ = 6 mb and $\sigma_
{J/\psi-M}^{abs}$ = 3 mb) and the model parameter $c_s$ is also reduced 
to 6.*10$^{-7}$.

In Fig. 4 the open circles with error bars indicate the preliminary
experimental data (cited directly from [41]) and the solid, dashed, dotted,
and dash - dotted curves are the theoretical results of the calculations
including the abs. + dis. + deg. (degradation) mechanisms, with abs. + dis.,
with abs. + deg., and with abs. only. One sees from Fig. 4 that all the four
curves are passing quite well through the pA and the AB (with light
projectile) data, however, only the solid curve can reproduce the Pb + Pb
datum point. The calculations with only absorption or the calculations with
abs. + deg. are not able to describe both, the Pb + Pb as well as
the other data points. However, the calculations with abs. + dis.
might be able to do so by an increase of the model parameter $c_s$ and the
effective cross sections as shown by the dashed curve in Fig. 1 .

In Figs. 5 and 6 the $J/\psi$ suppression factor is shown as a function of
the neutral transverse energy $E_T$ (GeV) for S + U reactions at 200 A GeV/c
and Pb + Pb reactions at 158 A GeV/c, respectively. The full squares
with error bars in Fig. 5 and the full squares and circles with error bars in
Fig. 6, respectively, refer to the preliminary NA38 and NA50 data
(cited directly from [26]). The different curves in these figures correspond
to the different calculations as explained in Fig. 4. Comparing Fig. 5 with 
Fig. 2 and Fig. 6 with Fig. 3 we find that the additional mechanism of 
leading particle energy degradation improves the description of the E$_T$ 
distribution of the $J/\psi$ suppression factor.

\section{Summation and acknowledgements}
In summary, we have introduced a mechanism for the effective dissociation of
$c\bar{c}$ pairs in the colour electric field of strings into a hadron and
string cascade model, i.e. JPCIAE [19]. This model is based on the LUND model 
and the PYTHIA event generator, simulating ultra - relativistic nucleus - 
nucleus collisions . The normal and the anomalous $J/\psi$ suppression in 
minimum bias pA, AB, and Pb + Pb collisions at 200 A GeV/c are qualitatively 
reproduced by this model relying on the mechanisms of the dissociation of 
$c\bar{c}$ states in the colour electric field of strings and of the nuclear 
absorption (both baryon and meson). However, the experimental data for the 
impact parameter (E$_T$) dependence of the $J/\psi$ suppression factor, both 
in S + U and Pb + Pb reactions at 200 A GeV/c and 158 A GeV/c, respectively, 
are not well reproduced. We also tested the additional mechanism of leading
particles energy degradation. Reducing the values of the effective cross 
sections and the model parameter $c_s$ we are able to reproduce the data of 
$J/\psi$ suppression factor vs. AB with the same quality as without this 
additional mechanism. However, the description of the E$_T$ distribution of 
the $J/\psi$ suppression factor is improved. 

Although it is not clear whether the environment of a $c\bar{c}$ state in the
colour electric field of strings should be regarded as dense hadronic matter
or as a QGP, the statement that any significant reduction of $J/\psi$ formation
in nuclear collisions would signal the creation of quark - gluon plasma has to
be revised [38, 46]. However, the origin of the dissociation of $c\bar{c}$
states in the colour electric field of strings should be studied further.

One of the authors, SBH, would like to thank the Institute of Theoretical
Physics, University of Tuebingen for the hospitality during his stay. We thank
also T. Sj\"{o}strand for detailed instructions of using PYTHIA. This work is
supported both by the NSFC of China and DFG of Germany as a joint scientific
research program. SBH would like to thank the Nuclear Industry Foundation of
China for financial support as well.

\bigskip
\bigskip
\bigskip
\bigskip
\newpage
\section*{Figure captions}
\begin{itemize}

\item[Fig.\,1:]
The $J/\psi$ suppression factor versus the product of the atomic mass
numbers of projectile and target nuclei, AB, in minimum bias pA and AB
collisions at 200 A GeV/c. The open circles with error bars are the
preliminary experimental data [4] taken directly from [41]. The dashed and
dash - dotted lines refer to the theoretical calculations including
absorption + dissociation (abs. + dis.) mechanisms and abs. only.

\item[Fig.\,2:] The $J/\psi$ suppression factor as a function of the neutral 
transverse energy in S + U reactions at 200 A GeV/c. The full squares with 
error bars are the preliminary NA38 data taken from [26]. The various curves 
refer to the different types of theoretical calculations as explained in Fig. 
1.

\item[Fig.\,3:] The $J/\psi$ suppression factor as a function of the neutral 
transverse energy in Pb + Pb reactions at 158 A GeV/c. The full squares and 
circles with error bars show the preliminary NA50 data taken from [26]. The 
various curves refer to the different types of theoretical calculations as 
explained in Fig. 1.

\item[Fig.\,4:] The $J/\psi$ suppression factor versus the product of the 
atomic mass numbers of projectile and target nuclei, AB, in minimum bias pA 
and AB collisions at 200 A GeV/c. The open circles with error bars are the
preliminary experimental data [4] taken directly from [41]. The solid, dashed,
dotted, and dash - dotted lines refer to the theoretical calculations including
absorption + degradation + dissociation (abs. + deg. + dis.) mechanisms, abs.
+ dis., abs. + deg., and abs. only.

\item[Fig.\,5:] The $J/\psi$ suppression factor as a function of the neutral 
transverse energy in S + U reactions at 200 A GeV/c. The full squares with 
error bars are the preliminary NA38 data taken from [26]. The various curves 
refer to the different types of theoretical calculations as explained in Fig. 
4.

\item[Fig.\,6:]The $J/\psi$ suppression factor as a function of the neutral 
transverse energy in Pb + Pb reactions at 158 A GeV/c. The full squares and 
circles with error bars show the preliminary NA50 data taken from [26]. The 
various curves refer to the different types of theoretical calculations as 
explained in Fig. 4.
\end{itemize}
\newpage
\begin{figure}[h]
\begin{center}
\leavevmode
\epsfxsize = 15cm
\epsffile[0 70 530 700]{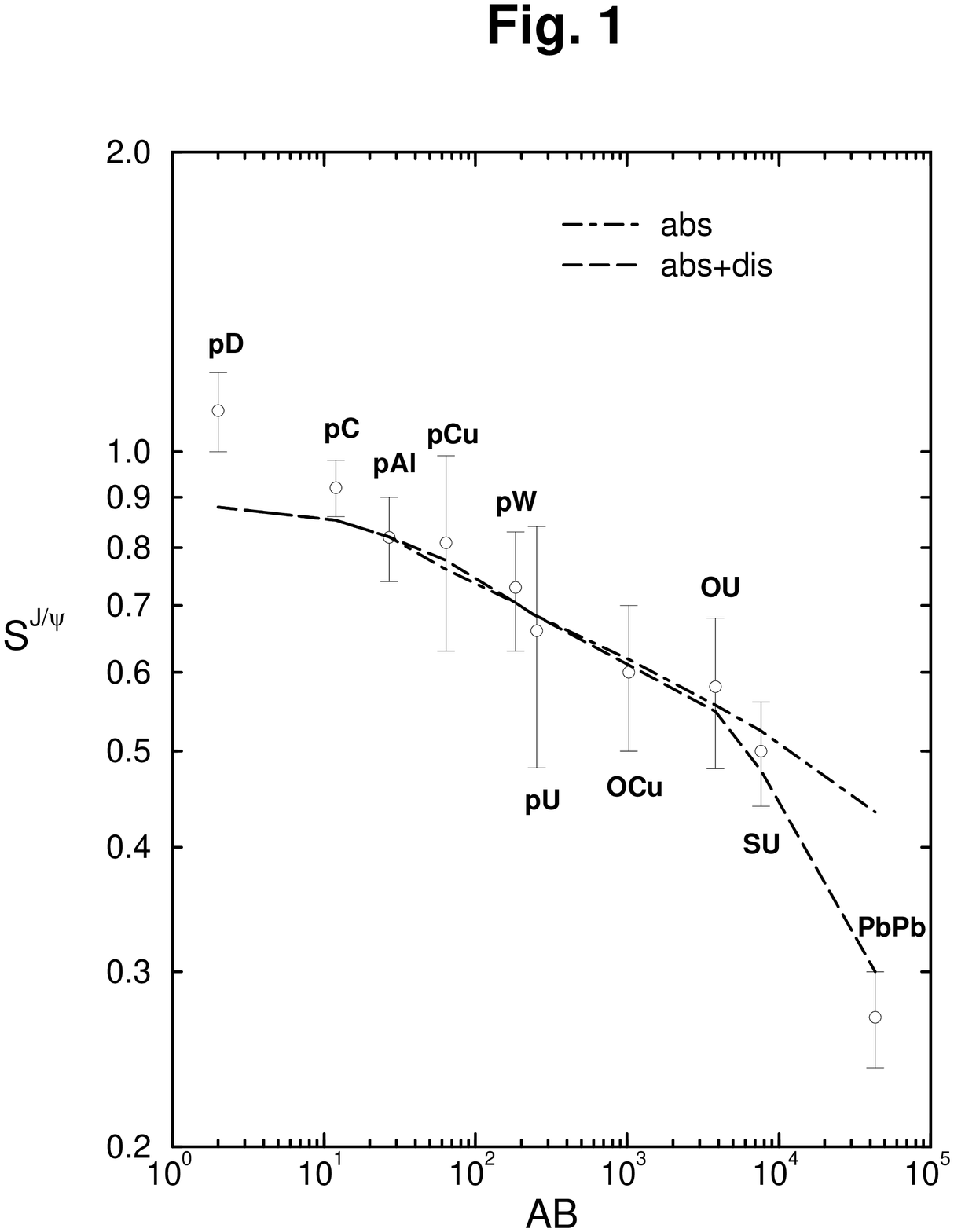}
\end{center}
\caption{}
\end{figure}
\newpage
\begin{figure}[h]
\begin{center}
\leavevmode
\epsfxsize = 18cm
\epsffile[0 0 730 740]{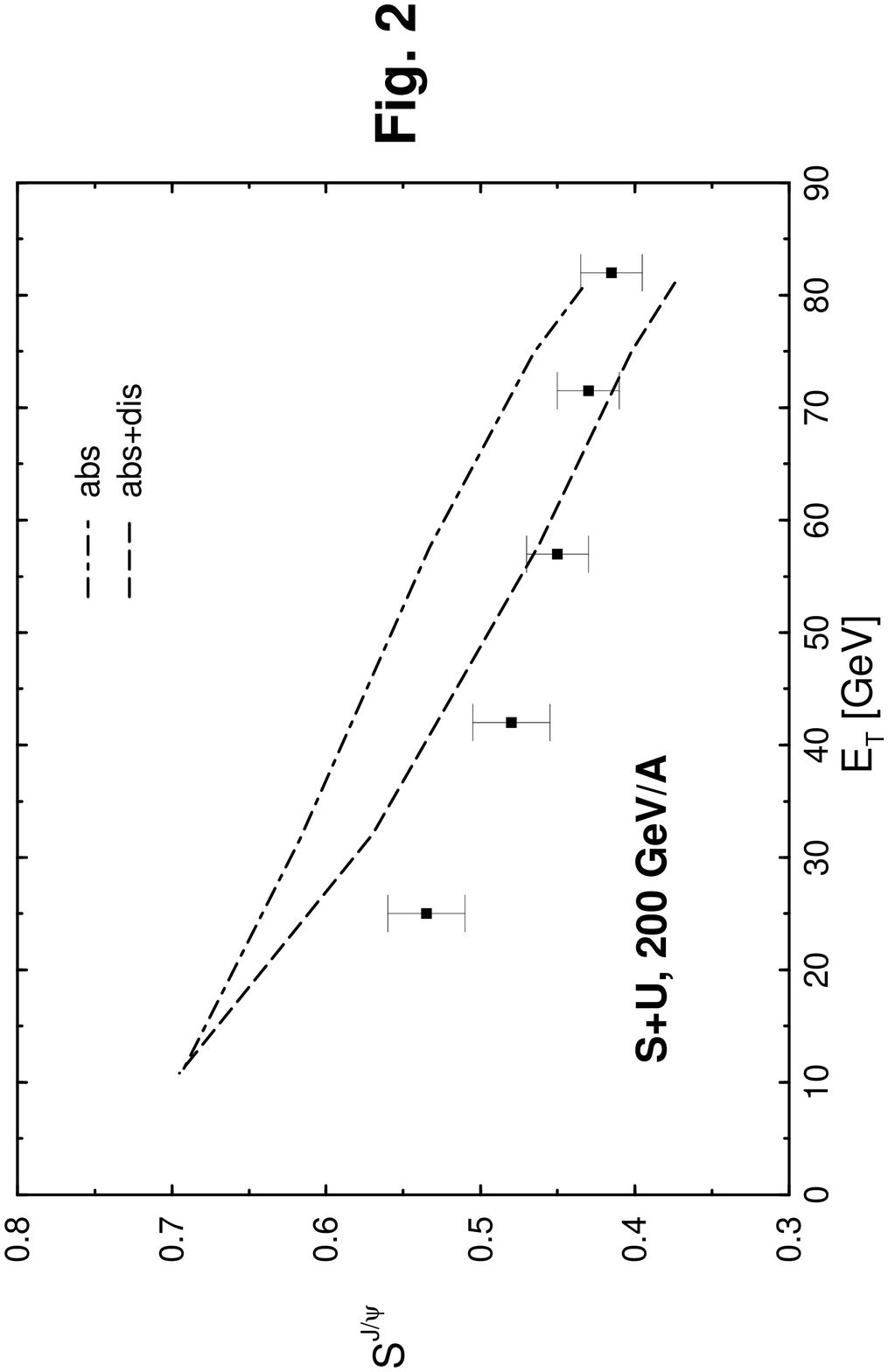}
\end{center}
\caption{}
\end{figure}
\newpage
\begin{figure}[h]
\begin{center}
\leavevmode
\epsfxsize = 18cm
\epsffile[0 90 730 740]{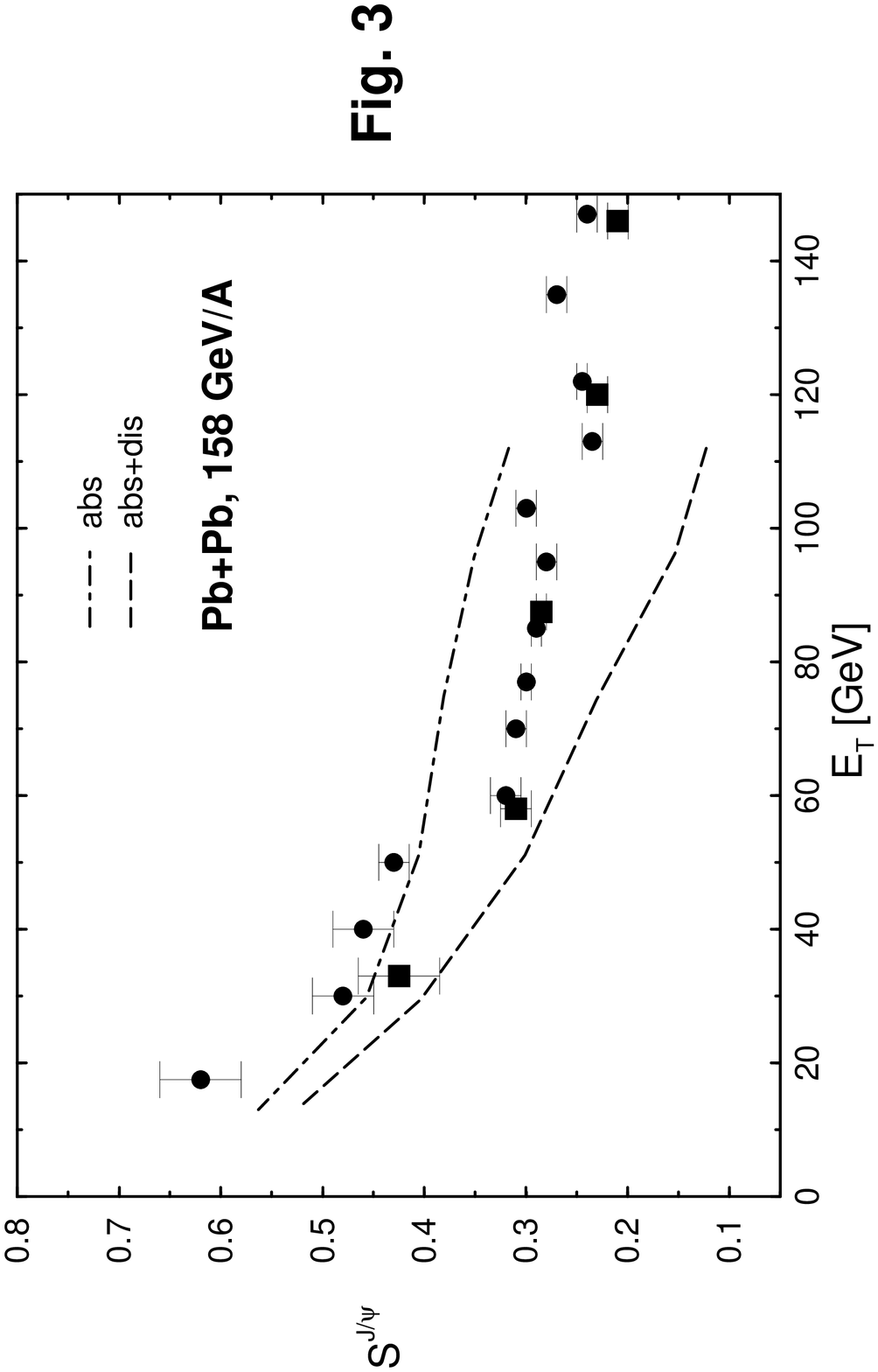}
\end{center}
\vspace{1.in}
\caption{}
\end{figure}
\newpage
\begin{figure}[h]
\begin{center}
\leavevmode
\epsfxsize = 15cm
\epsffile[0 70 530 700]{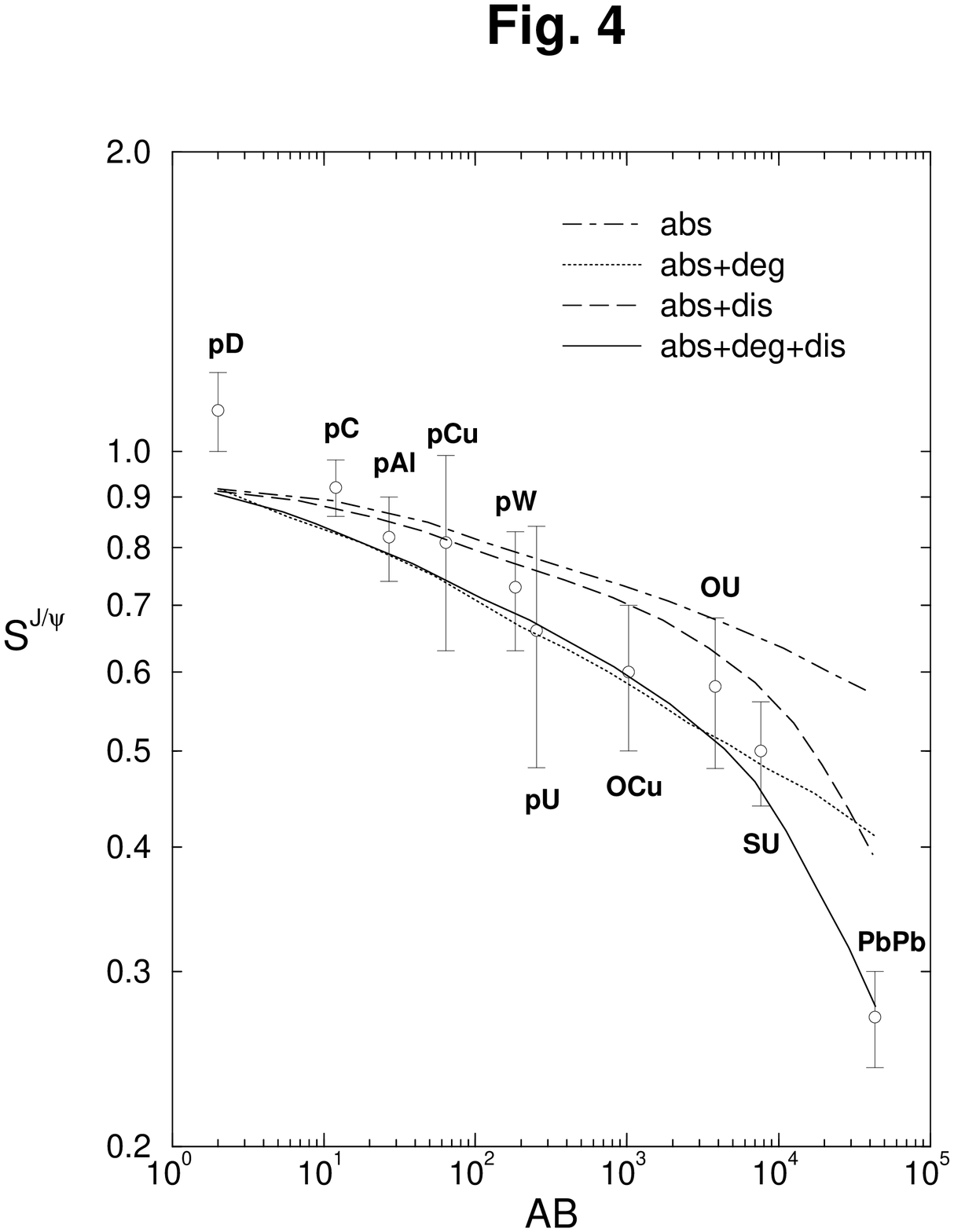}
\end{center}
\caption{}
\end{figure}
\newpage
\begin{figure}[h]
\begin{center}
\leavevmode
\epsfxsize = 18cm
\epsffile[0 0 730 740]{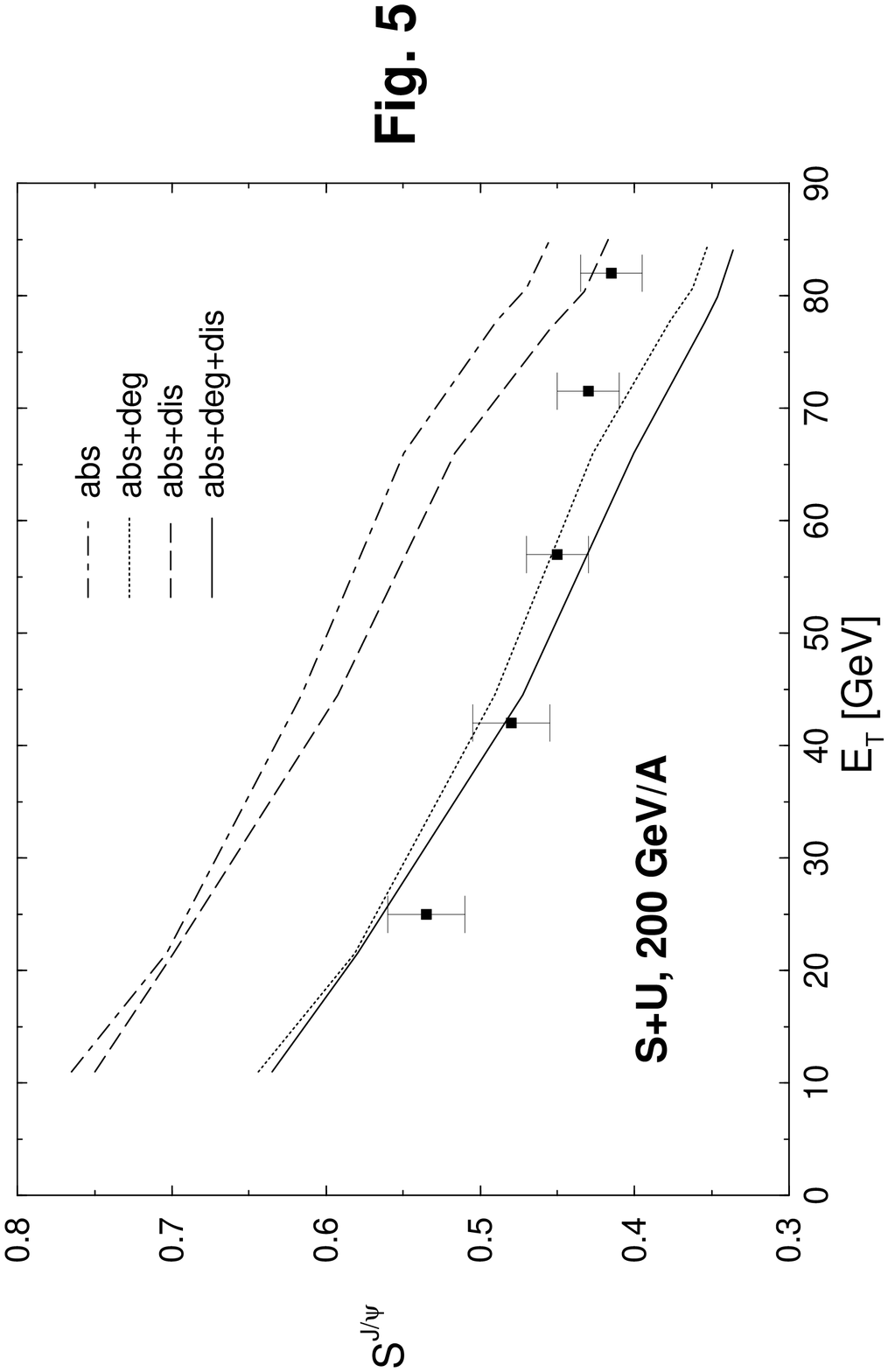}
\end{center}
\caption{}
\end{figure}
\newpage
\begin{figure}[h]
\begin{center}
\leavevmode
\epsfxsize = 18cm
\epsffile[0 0 730 740]{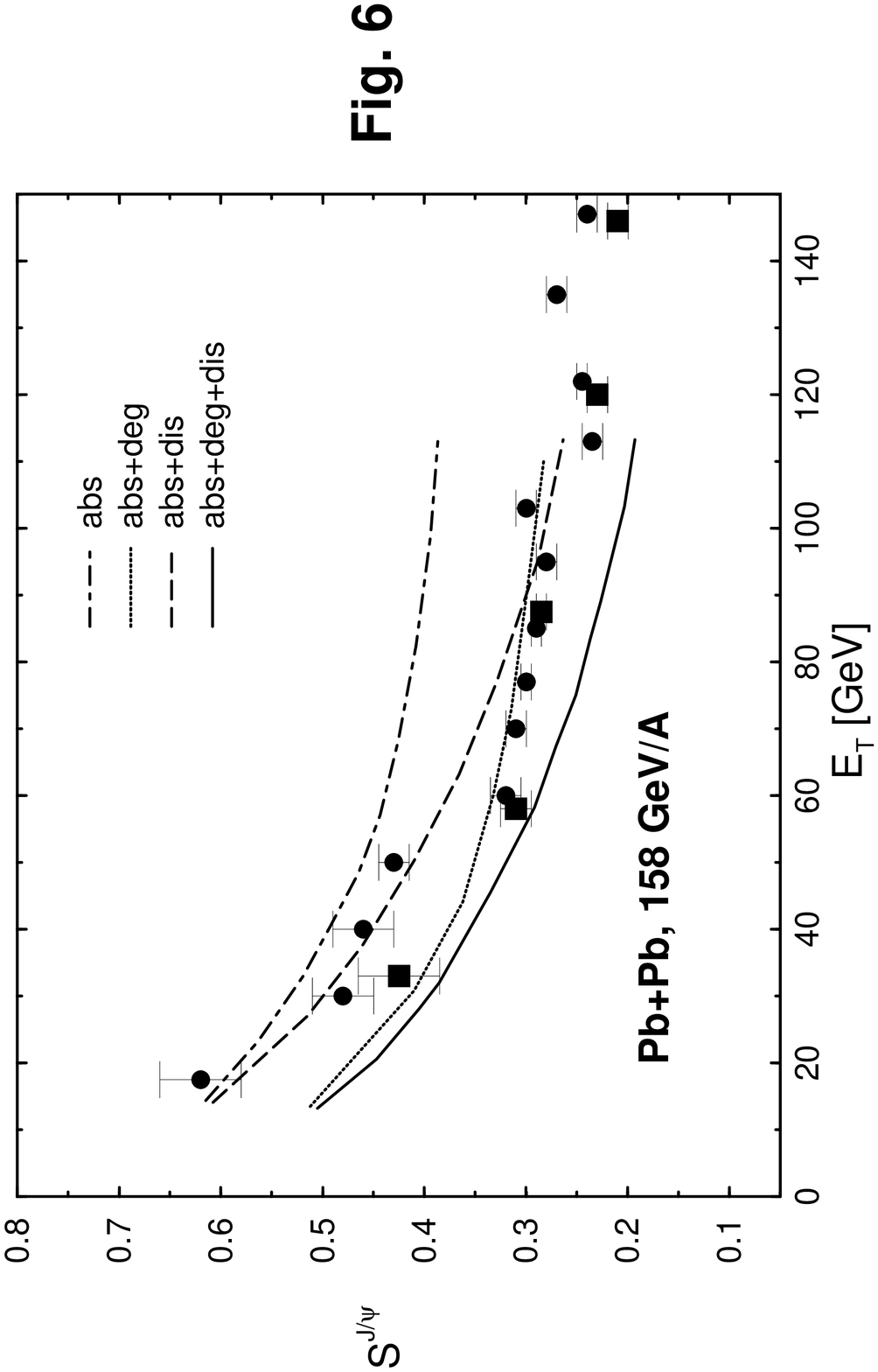}
\end{center}
\caption{}
\end{figure}

\end{document}